\def\year{2018}\relax
\documentclass[letterpaper]{article} 
\usepackage{aaai18}  
\usepackage{times}
\usepackage{helvet}
\usepackage{courier}
\usepackage{url}      
\usepackage{graphicx} 
\usepackage{amssymb}
\usepackage{amsmath}
\usepackage[ruled,vlined,algonl,boxed]{algorithm2e}
\usepackage{algorithmic}
\usepackage{enumitem}
\usepackage{colortbl}
\usepackage{microtype}

\newcommand{\stitle}[1]{\smallskip\noindent\emph{#1}\xspace}
\newcommand{\sstitle}[1]{\smallskip\noindent\textbf{#1}\xspace}
\newcommand{\ititle}[1]{\smallskip\noindent\emph{#1}\xspace}

\newcounter{prob}
\newtheorem{problem}[prob]{Problem}

\DeclareMathOperator*{\argmin}{arg\,min}

\title{Leveraging Quality Prediction Models for Automatic Writing Feedback}
\author{Hamed Nilforoshan {\normalfont and} Eugene Wu\\
Department of Computer Science\\
Columbia University\\
hn2284@columbia.edu, ewu@cs.columbia.edu\\
}

\begin{document}
\maketitle
\begin{abstract}
User-generated, multi-paragraph writing is pervasive and important in many social media platforms (i.e. Amazon reviews, AirBnB host profiles, etc). Ensuring high-quality content is important.  Unfortunately, content submitted by users is often not of high quality. Moreover, the characteristics that constitute high quality may even vary between domains in ways that users are unaware of.   Automated writing feedback has the potential to immediately point out and suggest improvements during the writing process. Most approaches, however, focus on syntax/phrasing, which is only one characteristic of high-quality content.

Existing research develops accurate quality prediction models. We propose combining these models with model explanation techniques to identify writing features that, if changed, will most improve the text quality. To this end, we develop a perturbation-based explanation method for a popular class of models called tree-ensembles.  Furthermore, we use a weak-supervision technique to adapt this method to generate feedback for specific text segments in addition to feedback for the entire document. Our user study finds that the perturbation-based approach, when combined with segment-specific feedback, can help improve writing quality on Amazon (review helpfulness) and Airbnb (host profile trustworthiness) by $> 14\%$ (3$\times$ improvement over recent automated feedback techniques).
\end{abstract}


 \begin{figure*}[t]
   \centering
   \includegraphics[height=4.75cm]{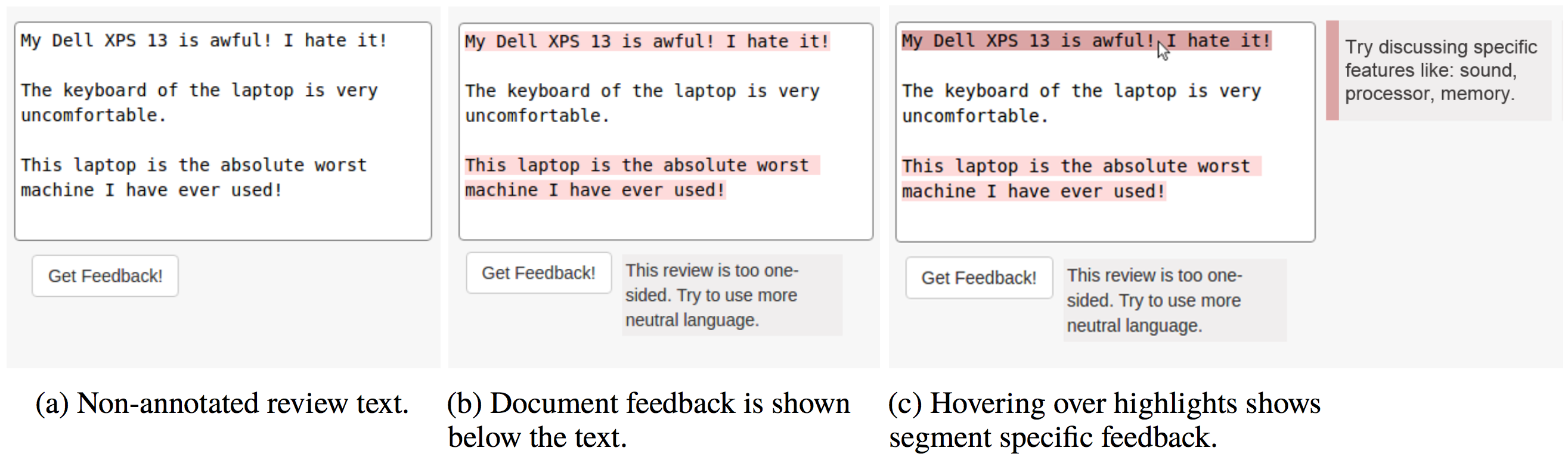}
   \caption{Example of feedback interface.}
   \label{f:screenshot}

 \end{figure*}


\section{Introduction}
\label{s:intro}

Modern social media relies on user-generated, multi-paragraph text. Product websites (e.g., Amazon, Yelp) rely on user-submitted product reviews to help users make better purchasing decisions; Q\&A services (e.g., StackOverflow, Quora) rely on the availability of user-generated questions and answers. A wide variety of examples fit this model, from profile descriptions in marketplaces (i.e Airbnb host profiles) to user comments on discussion boards (i.e reddit).

Content quality matters. Unfortunately, producing high-quality, written content can be challenging because the writing characteristics that indicate high quality often vary depending on the domain.  For instance, product reviews may prioritize comprehensiveness and specificity, whereas AirBnB host profiles prioritize trustworthiness\cite{amazonguidelines,ma2017self}.  Even well-intentioned writers may be unaware of these norms, which require substantial effort to learn and adhere to\cite{yelpguidelines,stackguidelines,wikiguidelines,amazonguidelines}.  

To this end, providing feedback during the writing process is valuable. However, providing online writing feedback at large scales and across different domains is a challenging problem. On one hand, crowd feedback techniques, both peer-based and payment based, can be effective at ensuring high-quality feedback~\cite{kulkarni2015,bernstein2010soylent}, but can be slow and costly to scale to the millions of internet users on most social platforms. On the other hand, fully-automated systems (i.e Grammarly and essay graders such as WriteToLearn) focus on grammar, misspellings, vocab, and other syntactic aspects of the writing. While such feedback can be helpful, it focuses on general English best-practices which is only a single, minor aspect of the local quality norms that are unique to each social platform\cite{yelpguidelines,stackguidelines,wikiguidelines,amazonguidelines}. For example, if an Amazon product review is eloquent and grammatically flawless, but provides no information about the product it reviews, such a review would be classified as low-quality. 

Our goal is to generate automated writing feedback that helps users adhere to a specific domain's norms for high-quality writing.   We observe an exciting opportunity: for a wide range of popular social media domains (e.g product reviews, profiles, comments), writing features which are predictive of domain-specific quality metrics have been identified. These features have been used to develop models which can accurately distinguish high or low quality in accordance with a platform-specific definition of quality~\cite{ghose2011estimating,cheng2014can,siersdorfer2010useful,ma2017computational,althoff2014ask}.  Furthermore, labeled datasets are publicly available to train these prediction models. However, while such models can predict writing quality, they do not provide suggestions on how to improve quality. 

In this paper, we propose the idea of using machine learning explanation techniques to generate specific suggestions for how to improve writing quality from existing quality prediction models. This is challenging because high-performing prediction models are often non-linear and difficult to translate to suggestions. Our primary insight is that, for a wide class of prediction models called tree ensemble models (e.g., random forests, gradient boosted trees, decision trees), which have been effective in a variety of text domains~\cite{ghose2011estimating,cheng2014can,siersdorfer2010useful}, it is possible to inspect their structure to learn how to improve the quality of a document.  To this end, we first introduce \textbf{Perturbation Analysis, a method for tree-ensembles, that identifies \emph{which} writing features (i.e tone, length) most contribute to the predicted label of ``low-quality'' for a document, and proposes \emph{how} to change the text to most improve the prediction} (Section~\ref{s:perturb}).

Perturbation Analysis relies on analysis of a pre-trained model; quality prediction models operate at the document level, so the suggested feedback will correspondingly be for the entire document. However, local feedback helps writers pin-point what writing segments to fix. Thus, we propose \textbf{using topic-based segmentation to split a document into topically cohesive text segments, and use weak supervision to train a segment-level model}. We then apply Perturbation Analysis to this model to generate per-segment feedback (Section~\ref{s:segmentation}).  Through a crowdsourced experiment, we show that this approach is feasible (Section~\ref{s:weak}).

We evaluate our methods on two real-life domains: Amazon Product Reviews and Airbnb Host Profiles (Section~\ref{s:eval}).  These domains use different quality definitions---review helpfulness and profile trustworthiness, respectively---and we show that our methods are effective across both domains.   We find that using Perturbation Analysis to provide localized feedback at the segment-level \textbf{improves the average submission quality by $>14\%$ across both domains, a $>3\times$ improvement on a state-of-the-art feedback system}. Perturbation Analysis produces statistically significant improvements over the state of the art for both Airbnb and Amazon, while localized feedback at the segment-level only does so when combined with Perturbation Analysis for the Amazon dataset~\footnote{https://github.com/cudbg/DialecticICWSM \\(Code and experiment data are available at this address)}.

\subsection{Approach Overview}
Figure~\ref{f:screenshot} illustrates our prototype feedback interface on a real product review.  After the user clicks ``Get Feedback!'', the system highlights low-quality segments in red in order to guide users. At first, only document-level feedback is shown (below the textbox) in order to not overwhelm the user.   Hovering over a highlighted segment shows the feedback on the side.  

\begin{figure}[b]
\centering
\includegraphics[width=.95\columnwidth]{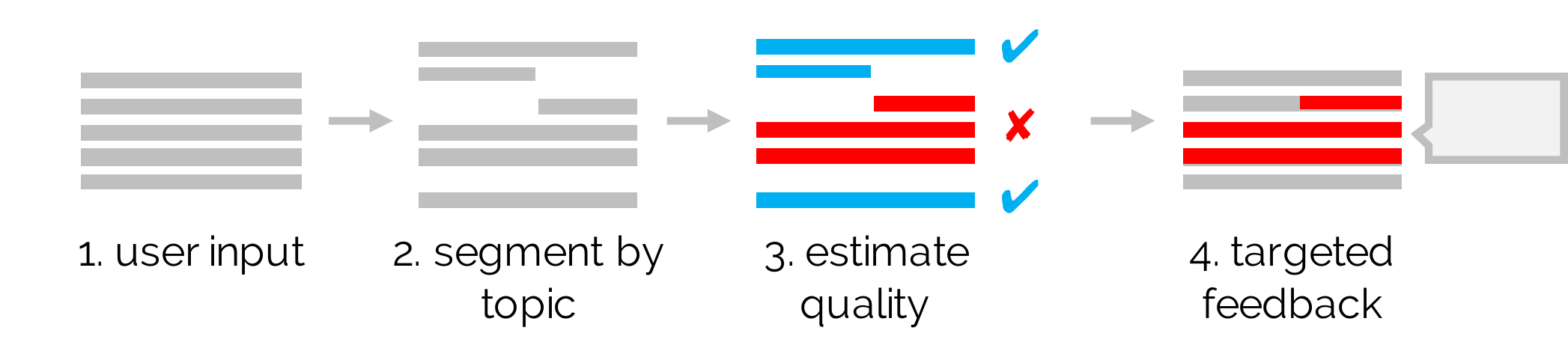}
\caption{Simplified process to generate segment feedback. }
\label{f:workflow}
\end{figure}

Figure~\ref{f:workflow} provides a general overview of our segment-level feedback generation.  The user submits their document (e.g., review), which we split into topic-cohesive segments (2). We then use pre-trained quality prediction models to estimate their quality (3), apply Perturbation Analysis to identify the writing features that should be improved (4), and show the targeted feedback to the user.  We also generate feedback for the entire document, which simply skips the segmentation step, and uses a prediction model trained for entire documents.

%
                                                          
\section{Related Work}

Our work is related to three main areas: (1)  existing approaches to writing feedback generation that our methods seek to improve upon; (2) social media text quality prediction models which we use to generate feedback via explanation; and (3) machine learning model explanation which inspire our novel feedback generation method. 

\sstitle{Generating Writing Feedback: } 
Collective intelligence can generate highly effective feedback. Learner-sourcing systems use peer-based feedback and scale to hundreds or thousands of students for writing and non-writing tasks~\cite{kulkarni2015,glassman2016learnersourcing}.  However, non-educational platforms are less likely to have peers willing to reliably write feedback.  Paid crowdsourcing such as Find-Fix-Verify~\cite{bernstein2010soylent} can also provide feedback or directly edit text.  However, human-based feedback can be prohibitively costly and slow for online communities (e.g., Amazon, Airbnb, reddit) with millions of participants.  

Automated approaches, on the other hand, are instant and can easily scale to large communities.  Automated essay grading has been widely used for assessing essays and writing exams.  However, essay graders either use complex prediction models to accurately grade written content but not provide feedback (e.g., GRE), or use simpler, interpretable models which can be used for feedback (e.g., linear regression with a small number of features) but are not accurate enough for many quality prediction tasks~\cite{farra2015scoring,valenti2003overview,hearst2000debate,attali2004automated}.  Our goal is to generate feedback {\it using more robust prediction models}. 

Other essay feedback tools (e.g., WriteToLearn and Grammarly~\cite{liu2016investigating,grammarly}) provide feedback on English best practices such as grammar, spelling, vocabulary use, structure, and phrasing.   However, these are not the only characteristics of high-quality social media content~\cite{yelpguidelines,stackguidelines,wikiguidelines,amazonguidelines}. For example, an eloquent and grammatically flawless product review devoid of any product information would be labeled low-quality (unhelpful). Vice-versa, a review that provides extensive information and commentary on a product with some mis-spellings would still be considered high-quality~\cite{ghose2011estimating}.  Effective social-media feedback systems should leverage writing features  predictive of the community's quality norms.

Towards this direction, Krause et al.~\cite{krause2015method} propose an outlier-based approach which can be extended to leverage features from existing text quality prediction models (including those for social media text).   They assume that the distribution of each feature follows a Gaussian distribution, and each feature's distribution is fitted using the feature's values from a training corpus of high-quality text.  For a given document, any outlier feature that is detected is mapped to a pre-written suggestion based on if the outlier should be increased or decreased to move closer to the distribution mean.  Note that although this method is extensible to a variety of features, in most cases an exceptionally high-quality document would also be marked as an outlier and be recommended to change in a way that results in lower quality. The limitations of outlier detection reveal why it is important for methods to take advantage of {\it state-of-the art prediction models}.

Finally, approaches that go beyond English best-practices only provide document-level feedback~\cite{krause2015method,farra2015scoring}.  Yet, research suggests that localized feedback is more valuable~\cite{nelson2007nature,kulkarni2015}.  Section~\ref{s:segmentation} studies segmentation methods for this purpose.

\sstitle{Social Text Quality Prediction Models: } 
There is a rich history of research which develops writing features in order to predict whether a user's writing adheres to an online community's quality standards. State-of-the-art models  predict \emph{community-specific notions of quality} including product reviews \emph{helpfulness}~\cite{ghose2011estimating}, Facebook/Twitter \emph{virality}~\cite{cheng2014can}, Youtube comment \emph{ratings}~\cite{siersdorfer2010useful}, Airbnb \emph{trustworthiness}~\cite{ma2017computational}, and reddit comment \emph{reactions}~\cite{althoff2014ask,tan2016winning}. We aim to analyze these prediction models in order to tailor writing feedback to a community's unique quality standards.

While these models can identify high or low-quality writing, they are not capable of producing feedback out-of-the-box. We observe that many of the aforementioned, state-of-the-art quality prediction models are tree-ensembles (e.g., random forests, gradient boosted trees, decision trees)~\cite{ghose2011estimating,cheng2014can,siersdorfer2010useful}.  Our Perturbation Analysis method probes a pre-trained tree-ensemble model to identify writing changes to a low-quality document that will most improve its predicted quality.

\sstitle{Interpreting Machine Learning: }
This area of research studies the relationship between machine learning model inputs and predictions.  
A common approach is based on sensitivity analysis, which treats the model as a black box, and studies how changes (perturubations) of each feature of an input affects its prediction~\cite{goldstein2015peeking,statslearn}.  Recent work computes  the ``impact score'' of a feature by measuring the model's sensitivity to changes in the feature's value, and visualizes the result to the user~\cite{krause2016interacting}.  Prior work studies each feature individually and does not account for multi-feature interactions that have been shown to influence model predictions~\cite{cortes1995support,breiman2001random}. For instance, if two features (i.e length and noun usage) have a positive effect on quality only when they both have high values, scoring them individually would result in underestimating their impact on model output.  

Our methods extend the idea of input perturbations to search for \emph{combinations} of  writing perturbations (edits) that will {\it increase} the model's predicted quality.  We find that exhaustively trying all possible perturbations is intractable, and develop heuristics that leverage the internal rule structure of tree ensemble models.

%
                                                            
\section{Perturbation Analysis}\label{s:perturb}

Perturbation Analysis rests on the idea that, assuming an accurate model, suggestions that increase the predicted quality are likely to increase the actual quality of the document.  Thus, as quality prediction models continue to improve, so too will the feedback that is generated.   We build on prior work that examines the sensitivity of the model prediction to small changes in individual features of a linear mode. Highly sensitive features are candidates for feedback~\cite{krause2016interacting}.

Prior techniques were designed for linear models.  Our work extends this idea to more complex tree-ensemble models, and accounts for multi-feature interactions (how combinations of changes simultaneousy affect a prediction). Tree-ensemble models are widely used in quality prediction~\cite{ghose2011estimating,cheng2014can,siersdorfer2010useful}, and include decision trees and random forests.  We describe our approach for an ensemble of trees, but the same method works for a single decision tree.

The rest of this section introduces our main idea using a toy example, introduces our formal problem statement that is independent of any specific model, and describes our heuristics that leverage tree model structures to quickly solve the feedback generation problem. Our evaluation will compare this approach with state-of-the-art outlier-based feedback generation~\cite{krause2015method}.

\begin{figure}[tbh]
  \centering
  \includegraphics[width=0.95\columnwidth]{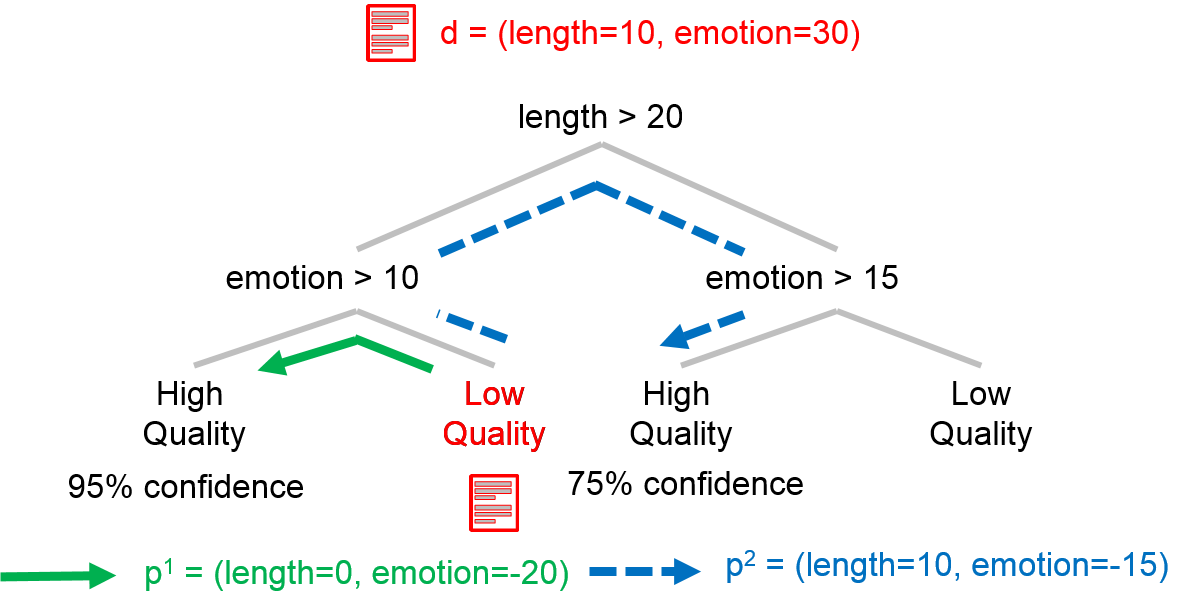}
  \caption{A simplified example to improve a document with features (length=10,emotion=30) that a decision tree has classified as low-quality. Applying {\it $p^1$} or {\it $p^2$} would re-classify the document as high-quality. Applying {\it $p^1$} to $d$ would decrease its emotion by $20$. Applying {\it $p^2$} to $d$ would increase the documents length by $10$ and decrease its emotion by $15$. }
  \label{f:tree}
\end{figure}

\sstitle{Simplified Example: } 
Figure~\ref{f:tree} shows a simplified example for a single decision tree.  $d$ is a document with two writing features: length and emotion, and is classified as low-quality.  The idea is to examine the minimum changes to $d$'s features so that it is classified in a high-quality leaf node.   These perturbations are shown as the green ($p^1$) and blue ($p^2$) paths.  Decreasing emotion by $20$ ($p^1$) leads to a  95\% confidence prediction; increasing the length by $10$ and reducing emotion by $15$ ($p^2$) leads to a $75\%$ confident prediction.   $p^1$ is preferable because it only modifies one feature (emotion) and has higher model confidence.   It is preferable to perturb fewer features so that the user needs to make fewer types of edits.  

Intuitively, a feature is more important if it is part of more paths that lead to a higher quality prediction, and it contributes more to a path if it contains fewer features that need to be changed.  To this end, we compute an ``impact score'' for each feature by aggregating its contribution to all paths that lead to a higher quality prediction.  In the example, emotion has a higher impact score because it appears in both perturbations ({\it $p^1$} and {\it $p^2$} ), and is the sole change for {\it $p^1$}.  This allows us to prioritize each feature based on the impact score.

\sstitle{General Problem Setup for Any Model: }
Let $\mathcal{F}$ be the set of $n$ model features, and $f_i$ denote the $i^{th}$ feature. 
Let $d \in \mathbb{R}^{n}$ be a data point (text document or segment) represented as a feature vector, where $d_i$ corresponds to the value of $f_i$.
For instance, $\mathcal{F}$ may be the set of text features described above, and a data point corresponds to the extracted text feature vector.
A model $M:\mathbb{R}^{n}\rightarrow\mathbb{R}$ maps a data point $d\in\mathbb{R}^n$ to a predicted quality score, where higher is better.  
A perturbation $p \in \mathbb{R}^{n}$ is a vector that modifies a data point, for a set of features. $p_i\ne 0$ if $f_i$ is a perturbed feature in the set, otherwise $p_i=0$.

Our goal is to identify feature subsets of the test data point $d$ that, if perturbed, will most improve $d$'s quality\footnote{We assume no feedback needed if data point is high quality.}.  
To do so, we first define the {\it impact} $I(d,p)$ for an individual perturbation $p$ as the amount that it improves the predicted quality discounted by the amount of the perturbation $\Delta(p)$ and the model's prediction confidence $C(d+p) \in [0, 1]$.  

$S^d_i$ computes the overall score for feature $f_i$ based on the impact of all perturbations involving $f_i$:
{\small
\begin{align*}
I(d,p) &= \frac{M(d+p) - M(d)}{\Delta(p)}\times C(d+p)\\
\Delta(p) &= \sum_{i \in [0,n]} dist(p_i)\\
S^d_i &= \sum_{p \in \mathbb{R}^{n}, p_i \ne 0} I(d, p)
\end{align*}
}
We compute $\Delta$ as the hamming distance, where the distance increases by one for each feature that is changed.  The confidence $C$ is the percentage of trees that vote for the perturbed prediction $M(d+p)$.

\begin{problem}[Feature Scoring Problem]\label{p:problem}
  Given $d$, return the top $k$ features in $\mathcal{F}$ with the highest overall scores $S^d_i$.
\end{problem}

\sstitle{Heuristic Solution for Tree Ensembles: }
The space of solutions for Problem~\ref{p:problem} is exponential in the number of features ($2^{n}$ perturbations for $n$ features). 
We present a heuristic whose complexity is linear in the number of paths in the ensemble of trees.
The idea is to compute perturbations and scores for each tree individually, and then aggregate their scores.
Below, we walk through the procedure for the tree in Figure~\ref{f:tree}, and refer readers to the Appendix (Section 9.1) for full mathematical details.

Let the tree in Figure~\ref{f:tree} be tree $T_i$ in the ensemble model $M=\{T_1,\dots,T_t\}$, and $D$ be the dataset.  Each leaf matches $l$ points in the dataset, and the leaf's quality label corresponds to the majority label of the $l$ points.  For instance, $q = (len\le 20, emotion\le 10)$ corresponds to a leaf node where 95\% of the matching points vote ``high quality''.   The minimum perturbation so that $d$ satisfies $q$ is to reduce emotion by $20$.    Thus, rather than trying all possible perturbation (e.g., reduce emotion by $30$), we simply examine the minimum perturbation for each path that leads to a ``high quality'' prediction.  The confidence $C$ for a given perturbation is the confidence of the corresponding path.   The score $S^d_i$ is computed as the total impact for all minimum perturbations across all paths in all trees.

\sstitle{Implementation:} 
This procedure is also very efficient because we can index all paths in the tree ensemble by their quality, and precompute the confidence for each path.  The index lets us retreive paths that increase the predicted quality, and in parallel, we can estimate the perturbation distance to compute impact scores.   Finally, we add the impact for each $p$ to the score for each feature in $p$.   We then rank the features by their total scores.  Our implementation generates feedback in $\approx\frac{1}{10}$ seconds.  

\ititle{Normalization: } 
We find that features closer to the root will happen to occur in more feature sets and have artificially higher scores.  Although being close to the root signals importance {\it for the average document}, for any specific document, we want to treat each feature equally. We adjust feature impact scores to reduce bias by drawing a sample of low quality text from the corpus.  For each feature $f_i$, we compute the sample sample mean $\mu_{f_i}$ and standard deviation $\sigma_{f_i}$ of impact scores, and normalize a feature's score $S^d_{f_i}$ by computing $\frac{S^d_{f_i} - \mu_{f_i}}{\sigma_{f_i}}$. 

\ititle{Feature Suggestion Direction: }
Finally, we compute the net sign across all observed perturbations to suggest whether to increase or decrease the value of the feature.

%
%

\section{Segment-Based Feedback}\label{s:segmentation}

While feedback based on syntactic, grammatical or phrasing mistakes can pin-point where the specific error occurs, more general feedback approaches typically offer suggestions at the level of the entire document~\cite{krause2015method,valenti2003overview,hearst2000debate,attali2004automated}.  Yet feedback psychology suggests that localized comments are more helpful for improvement~\cite{nelson2009nature,kulkarni2015}.  Our approach is to split the document into smaller segments and provide feedback on a per-segment basis.  There are two challenges: how to perform the segmentation, and how to get around the issue that we do not have training labels for individual segments.

\subsection{Segmentation} 
Psychology research suggests that mentally processing the topical hierarchy of text is fundamental to the reading process~\cite{hyona2002individual}, and
contributor rubrics across many social media services suggest that content be structured by topics~\cite{yelpguidelines,amazonguidelines,wikiguidelines}.
For this reason, we propose segmenting input documents by topic.  We believe this is preferable to segmenting by sentence, which would result in very short segments, or by paragraph, because not all writing requires a multi-paragraph form.  

For this purpose, we use TopicTiling~\cite{riedl2012topictiling}, an extension to TextTiling~\cite{hearst1997texttiling}, that uses a sliding window approach.  It  computes the LDA topic distribution within each window of text and creates a new segment when the distribution changes beyond a threshold.  
TopicTiling is also competitive with other similar segmentation algorithms~\cite{riedl2012topictiling}.
It is important to note that generating segment-based feedback is not fundamentally dependent on any particular algorithm, and developers may employ their preferred segmentation algorithms.

\subsection{Segment-level Prediction Model}\label{s:weak}

A prediction model trained to predict document-level quality is ill-suited for individual segments. Consider a review that is excessively long, but each segment is short and lacks content.  We may want to suggest that the writer reduce the overall review's length, but ask to elaborate further for specific segments.  The challenge is that existing quality labels are at the granularity of the document rather than at the granularity any specific segmentation algorithm's output.  To this end, we employ weak supervision by assuming that document quality is sufficiently correlated with each segment's quality, and a document's label can be used to label its segments as training data for a segment classifier. In other words, for training purposes, we assume high-quality documents are composed of high-quality segments, and low-quality documents are composed of low-quality segments.  This general approach has been effective in other machine learning domains~\cite{banovic2017leveraging}.  

\stitle{Validation of Weak Supervision: }
We evaluate our weak supervision assumption using the Amazon review helpfulness corpus by conducting a crowdsourced experiment\footnote{Full details described in Appendix (Section 9.2).}. We first segment each document in the Amazon review corpus by topic, and train a quality prediction classifier on a training set of segments labeled by their document-level quality labels, using a random forest model, with features and document labels consistent with our later experiments (Section~\ref{s:custom}). We then evaluate the accuracy of this model on a similarly labeled test set of segments, and compare its accuracy with the same test set labeled by humans.  Comparable accuracy suggests evidence to validate the weak supervision approach.  The classifier first achieves $72.5\%$ accuracy at predicting the weak, document-level labels, on a balanced, test sample of $500$ segments (from $250$ helpful \& unhelpful reviews).  

We then crowdsourced human labels for the same test sample of $500$ segments. We train crowd workers by showing them a separate sample of segments, along with explanations of why each segments is helpful or unhelpful. We then randomly assign each worker 50 segments to label, collected $\ge 3$ labels per segment, and determined the final label of each segment using the Get Another Label algorithm~\cite{sheng2008get}. The classifier trained under weak-supervision predicted these human labels with $71.1\%$ accuracy.   We deemed this as promising results,  although more studies are needed to fully evaluate this hypothesis across other text domains. For completeness, we report that the  model trained via weak supervision for segments in the AirBnB dataset achieved $69.3\%$ accuracy on the document-level labels (not crowdsourced) of a balanced test-set of 300 segments.  The accuracy was similar enough to the Amazon corpus that we did not rerun the crowdsourcing component.

%
%

\section{Evaluation}
\label{s:eval}

We evaluated our feedback methods on two datasets, Amazon product reviews and Airbnb host profiles~\cite{amazondataset,ma2017computational}, which employ different quality measures: review helpfulness~\cite{ghose2011estimating} and profile trustworthiness~\cite{ma2017self}, respectively. We chose these datasets because they are publicly available, exemplify multi-paragraph writing, and quality prediction models have already been developed for them by prior work. We sought to test two hypotheses:

\begin{itemize}
\item (H1) Feedback generated using Perturbation Analysis is more effective that feedback generated by state-of-the-art methods that leverage writing features, and 
\item (H2) supplementing document-level feedback with localized (Segment-based) feedback can further improve writing quality.  
\end{itemize}

\begin{table*}[ht]
\hspace*{-.1in}
\centering
\small
\begin{tabular}{m{6.25em}|m{1em}|m{25em}|m{17.5em}}
    {\textbf{Category}} &  {\textbf{\#}} & {\textbf{Summary of Model Features}} & {\textbf{Description of Feedback}}  \\ \hline
    Readability and Grammar & 14  &  ARI, Gunning index, Coleman-Liau index, Flesch Reading tests, SMOG, punctuation, parts of speech distribution, lexical diversity measures, \it{LIWC grammar features} & To revise writing style to be more clear 
 \\ \hline

    \textbf{Subjectivity}  & 15 &  \textbf{opinion sentence counts~\cite{Hu2004}, valence, polarity, and subjectivity scores and distribution across sentences ~\cite{ghose2011estimating,gilbert2014vader,loria_2014}, \% upper case chars, first person usage, adjectives } & \textbf{To revise writing to be more balanced, while preserving opinions about product} \\ \hline

    Informativeness  & 8 &  mined jargon word (i.e product features for Amazon) and named entity stats~\cite{Hu2004,Agrawal1994}, length measures (word, sentence, etc. count) & To go into more detail (recommends jargon, mined using~\cite{Agrawal1994})  \\ \hline

    Topic & 5 &  LDA topic distribution and top topics ~\cite{blei2003latent}, entropy across topic distribution & To reduce irrelevant topics, focus on product/host-related topics (recommends topics~\cite{blei2003latent})  \\ \hline

    \it{Friendliness} & 5 &  \it{LIWC friendliness, social, we, family, inclusive measures} & \it{To add additional friendly/social text.}  \\ \hline


    \end{tabular} 

\caption{Summary of writing features by (left to right): the category name, the number of features, a summary of the features, and a description of the feedback text. \textbf{Bold} and {\it Italics} indicate features unique to Amazon and Airbnb domains, respectively.}
  \label{t:features}
\end{table*}

\subsection{Customizing to Application Domains}\label{s:custom}

This subsection describes how model training and our feedback text were customized for the Amazon and Airbnb domains. 
The overall setup was to use pre-existing text features developed by existing quality prediction research to train the document-level and segment-level models, and to follow prior approaches that map features to pre-written feedback text~\cite{krause2016interacting,krause2015method}.  

\subsubsection{Model Training}
We trained  different document- and segment-level models for each domain, using the same features.  Both datasets encode continuous quality measures.  Amazon helpfulness consists of the fraction of users that labeled the review ``Helpful'' (as opposed to ``Not Helpful''), whereas Airbnb trustworthiness consists of a crowdsourced continuous score.  Following prior work, we transform these into binary classification models by defining thresholds for ``high-quality'': $\ge60\%$ ``Helpful'' for Amazon reviews~\cite{ghose2011estimating} and $\ge$median for Airbnb trustworthiness~\cite{ma2017computational}.  

We do not innovate on developing new features and simply borrow those that exist in the literature. For Amazon, we start with the Readability/Grammar, Informativeness, and Subjectivity feature categories developed by Liu et al.~\cite{liu2007low}. We then augment these  with additional features found in other review helpfulness prediction literature~\cite{kim2006,chen2011quality,chen2011quality,liu2008,ghose2011estimating}. Lastly, we add a Topic category which uses LDA features, which have been demonstrated to perform well at review classification~\cite{blei2003latent}. For Airbnb, we retain the Readability/Grammar, Informativeness, and Topic categories based on prior work which shows that features relating to each of these three categories is predictive of trustworthiness~\cite{ma2017computational,ma2017self}. Readability/Grammar is augmented with all LIWC features identified by Ma et al. that relate to Grammar. We also create a fourth category, Friendliness, to group similar LIWC features identified as predictive by Ma et al.~\cite{ma2017computational}. Each model thus used 4 feature categories (\textbf{Subjectivity} unique to Amazon, {\it Friendliness} unique to Airbnb); categories consist of between 4 to 15 feature extractors.  

Finally, we used a random forest model, similar to existing work. Our trained document-level classifier performs similarly to or better than state-of-the-art models.  For Amazon, we achieve a similar $85\%$ accuracy on a balanced sample of 500 reviews~\cite{ghose2011estimating}.    For Airbnb, we achieve $79.3\%$ accuracy on a balanced sample of 300 profiles, as compared to $68\%$ in prior work~\cite{ma2017computational}; slight additions in feature set (i.e word count) explain the improvement.

\subsubsection{Generating Text Feedback } 
Given a top set of features and the direction they should change (increase/decrease) from Perturbation Analysis, we want to translate them into textual feedback.  Prior work simply maps each feature to pre-written text---for example, a sentiment feature would be mapped to the text ``Please try to make your writing more/less positive'' depending on the suggested direction~\cite{krause2015method}.  We take a similar approach with two slight variations.

Firstly, we note that giving feedback for each individual feature in our models would be redundant. Consider the Readability category of features where multiple readability measures are (i.e ARI, SMOG, Flesch-Kincade, etc) are a part of the model. Feedback for each readability feature would effectively be the same (to improve the clarity of the text). Similarly, the feedback for low \emph{Informativeness} is to suggest writing more and to go into more detail. We note that the same issue of feedback redundancy applies to the features of each category. Thus, we instead compute a category impact score (the average across all its constituent features) and show feedback only for the highest category score. The specific text for each suggestion was based on prior work that highlights the primary reasons why Amazon reviews are labeled not helpful and Airbnb profiles are labeled not trustworthy~\cite{badreviewreasons2,ma2017self}.

Finally, in some cases it is helpful to return dynamic feedback based on the document.  For instance, the topic related features use LDA to identify topics (e.g., ``Durability'' for product reviews or ``Family'' in profiles) that are in the document.  This could be leveraged to also suggest missing topics to include if Perturbation Analysis finds that the document is missing topics. We use this approach for feedback on the Topic category, using topic labels by Ma et al. for Airbnb~\cite{ma2017computational}, and manual labels for the 15 LDA topics from Amazon. In cases where manual labeling of topics is not possible, automated approaches can also be used~\cite{mei2007automatic,ritter2012open}. We take a similar approach for the Informativeness category, using the apriori algorithm to recommend specific jargon (e.g., ``price'' for product reviews, ``outdoors'' for profiles) for writers to discuss~\cite{Agrawal1994}.

Note that the feedback could be further customized based on the magnitude of the suggested perturbations, the specific features, or even using alternative feedback generation methods.  For instance, essay grading methods mainly relate to Readability, and could be used to provide more detailed grammatical or stylistic feedback.  We found that our simple methods worked well, and leave the problem of text customization to future work.

\subsection{Experiment Design}

We used a crowdsourced study on Mechanical Turk that compared four feedback systems along two dimensions.  {\it Granularity} compares document level ({\it Doc}) with document {\it and} segment level ({\it Seg}) feedback.  {\it Explanation Selection} compares the outlier-based feedback~\cite{krause2015method} ({\it Krause}) with Perturbation Analysis feedback ({\it Perturb}).  We used {\it Krause} as the baseline because it is recent, and also proposed to leverage model features for feedback generation.    This results in a 2x2 between-subjects design, where each participant was randomly assigned to one of four conditions: \emph{Doc+Krause}, \emph{Seg+Krause}, \emph{Doc+Perturb} and \emph{Seg+Perturb}.

{\it Krause} is based on an outlier-based feedback method~\cite{krause2015method} described in the Related Work.  It was shown to out-perform static writing rubrics in the context of university peer-based code reviews.  The summary is that it computes the mean and standard deviation of each writing feature based on a sample of high quality documents (we also adapted it to segments for the localized feedback setting).  If a document's feature is $\ge1.5$ standard deviations from its mean, the pre-written feedback text is shown for that feature.   Our implementation extends Krause et al.~\cite{krause2015method} with application specific features that were given high weight in our random forest models.  This is so that the comparison results are due to feedback mechanisms, rather than the specific model features.    To this end, we added informativeness and readability features (the \# of product features/jargon, and Coleman-Liau index) for the Amazon experiment, and self-disclosure and friendliness features (\# topics and LIWC socialness measures) for the Airbnb experiment.

\stitle{Amazon Participants:} We recruited $85$ workers on Amazon's Mechanical Turk (38.8\% female, ages 20-65 $\mu_{age}$=32, $\sigma_{age}$=8.5). $81$ workers completed the task. Participants were randomly assigned to one condition group; all conditions had 21 subjects except the \emph{(Seg+Perturb)} condition which had 22.  No participant had used our feedback systems before. $71.3\%$ had written a prior product review; all had read a product review in the past. All participants were US Residents with $> 90\%$ HIT accept rates. The average task completion time was 14 minutes, and payment was $\$2.5$ ($\sim\$10/hr$).

\stitle{Airbnb Participants:} We recruited $92$ workers (41.3\% female, ages 20-62 $\mu_{age}$=33, $\sigma_{age}$=8.2) on AMT. 91 workers completed the task. 21, 26, 22, and 23 participants were were randomly assigned to conditions 1, 2, 3, and 4 respectively.  We used the same HIT qualifications as the Amazon experiment. $62\%$ of participants had used AirBnb before. The average completion time was 11 minutes, and payment was $\$2.5$ ($\sim\$13.6/hr$).

\stitle{Procedure: } 
Participants writing product reviews were asked to write a {\it review of their most recently owned laptop computer} ``as if they are trying to help someone else decide whether or not to buy that laptop, and are writing on a review website like the Amazon store''. We used a qualification task to ensure participants had ever owned a laptop. Participants writing Airbnb profiles were asked to ``pretend that [they] are interested in being a host on Airbnb'' and to ``write an Airbnb profile for [themselves]''. Participants were told that, upon submitting their writing, they {\it may} receive feedback and could {\it optionally} revise.

Upon pressing the {\it I'm Done Writing} button, the interface displayed document-level feedback under the text field; for users in the segmentation condition, low quality segments were highlighted red and the related feedback displayed when users hovered over the segment (Figure~\ref{f:screenshot}). We then gave participants the opportunity to revise their submission; to avoid bias, we clearly state that they were ${\textbf{not}}$ obligated to revise their text.   Users could click the {\it Recompute Text Feedback} button to see updated feedback on their revisions (median 1 click/participant), or submit and finish the task. We used a post-study survey to collect demographic information as well as their subjective experience.

The interface was the same for all conditions---only the feedback content changed. The final submission was considered the {\it post-feedback} submission, and the initial submission upon pressing the {\it I'm Done Writing} was the {\it pre-feedback} submission.  The experiment was IRB approved.

\subsection{Quality Scoring}

$81/85$ participants completed the product review writing task, and $91/92$ completed the host profile writing task. 
A panel of three expert evaluators (non-authors), recruited from the researcher's university, coded the pre {\it and} post-feedback documents using a rubric based on prior work on review quality~\cite{badreviewreasons2,liu2007low} and Airbnb profile quality ~\cite{ma2017self}.

Both rubrics defined quality for their respective domains, and provided examples of low/high-quality documents. The Amazon rubric described review quality as helpfulness to potential laptop shoppers---based on the three helpfulness subcomponents of Informativity, Readability, and Subjectivity, identified by prior work. The rubric explained the meaning of each subcomponent and connected them to overall helpfulness~\cite{badreviewreasons2,liu2007low}. The Airbnb rubric defined quality as trustworthiness for potential tenants---based on the three components of trustworthiness identified by \cite{ma2017self}: Ability, Benevolence, and Integrity. After asking evaluators to assess documents on each subcomponent, the rubric asked them to rate the holistic quality of the documents on a 1-7 point scale (taking into account the three respective subcomponents and any other salient aspects of the document). We use the change in this score between pre and post-feedback to measure feedback utility. 

Each score is the average rating from two coders---if they differed by $\geq 3$, the third coder was used as the tie breaker and decided the final value.  We trained the third coder by showing them the Amazon or Airbnb corpus, examples across the quality spectrum, and the disputed ratings from the other two coders.  The coders labeled documents in random order and did not have access to any other information about the documents.

\subsection{Quantitative Results and Analysis}

\begin{figure}[t]
\centering
\includegraphics[width=0.95\columnwidth]{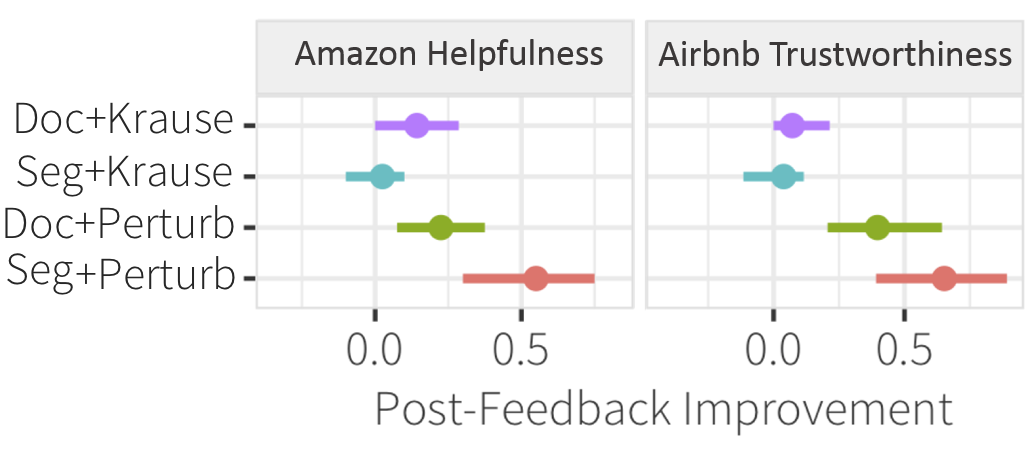}
\caption{Overall Quality improvement scores across all four conditions, for Airbnb profiles and Amazon product reviews}
\label{f:improvement}
\end{figure}

Figure~\ref{f:improvement} plots the mean change and $95\%$ bootstrap confidence interval for the four conditions scores, for both text domains. \textbf{These plots show that the largest improvements were achieved by using \emph {Perturbation Analysis} to provide \emph{localized} (Segment-based) feedback.} For Amazon product reviews, {\it Seg+Perturb} improved overall helpfulness by over $3.9\times$ over the baseline {\it Doc+Krause} (0.55 vs. 0.14 increase), and had a $2.4\times$ improvement over the next-best {\it Doc+Perturb} condition (0.55 vs. 0.22).  Similarly, for Airbnb profiles, {\it Seg+Perturb}, improved the overall trustworthiness by over $9.1\times$ over the baseline {\it Doc+Krause} (0.65 vs. 0.07 increase), and had a $1.6\times$ improvement over the next-best {\it Doc+Perturb} condition (0.65 vs. 0.40 increase).

\sstitle{Are Perturbation Analysis and Segment-based Feedback Effective?} 
We then performed a two-way ANOVA using overall quality increase as the dependent variable, and Perturbation Analysis ({\it Perturb}) and Segment-based Feedback ({\it Seg}) as the independent variables, for both Airbnb and Amazon. The ANOVA found a significant effect for {\it Perturb} for both Amazon helpfulness (F(1,80)=9.66, p=0.0026$<$.005) and Airbnb (F(1,90)=26.34, p$<$.0001). We can see this in Figure~\ref{f:improvement}: {\it Seg+Perturb} outperformed {\it Seg+Krause}, and {\it Doc+Perturb} outperformed {\it Doc+Krause}.  On the other hand, {\it Seg} alone did not have a significant effect on Amazon (F(1,80)=2.21, p=0.14) or Airbnb (F(1,90)=1.14, p=0.29).  Finally, interaction effects between {\it Perturb} \& {\it Seg} were significant for Amazon (F(1,80)=5.75, p=0.019$<$.05) but not Airbnb (F(1,90)=2.69, p=0.10). Moreover, we found that for both Airbnb and Amazon there was no significant difference between the \# of feedback messages shown at the document-level (two-way ANOVA) or at the segment-level (one-way ANOVA) for any of the conditions. We therefore believe that the feedback method and content, not the frequency (sensitivity), was what most influenced the outcomes. \textbf{These results suggest that Perturbation Analysis clearly outperforms the baseline in generating feedback (H1), but there is no evidence that localized (Segment-based) feedback, in isolation, necessarily improves document-level feedback (H2)}. Significant interaction effects for Amazon, however, reveal that it is necessary to further investigate the effectiveness of localized feedback.

\sstitle{When Does Segment-Based Feedback Work?} s
Given the interaction effects in the Amazon experiment, we look more closely to see if Segment-based feedback produces significant improvements in any condition. Using Tukey's HSD post-hoc test to compare the individual conditions for Amazon product reviews, we found that the pairwise comparisons between {\it Seg+Perturb} and all other conditions were significant (p$<$.05). However, all pairwise comparisons between the latter three conditions did not show significance. Thus, Segment-based feedback did significantly improve feedback quality in the case of Amazon, but only when providing feedback via Perturbation Analysis. This suggests a co-dependency between segmentation and perturbation methods. We infer that Segment-based feedback may only be useful when the suggestions are generated using a robust method (i.e Perturbation Analysis); using simple outlier detection suggestion to generate Segment-based feedback will not produce helpful feedback. 

However, as mentioned earlier, segment-based feedback did not achieve a similar outcome in the Airbnb experiment and there were no interaction effects. We suspect the difference in improvement for the {\it Seg+Perturb} condition between the two domains is because the Amazon reviews written by participants contained 1.7 more segments on average as compared to the profiles of Airbnb participants ($t(171)$=7.77, p$<$0.0001, $\mu_{amazon}$=$4.1$, $\mu_{airbnb}$=$2.7$), and thus in the case of Airbnb, there was simply less writing to provide localized feedback on. 

\textbf{In summary, evidence suggests that H2 (Segment-Based Feedback efficacy) was only true when providing localized, Perturbation-Based feedback for Amazon, where there were more segments on average to provide feedback for.}

\subsection{Qualitative Results}

\sstitle{Strengths of Our Methods:}
In the post-experiment feedback form, many users (N=8) in the {\it Perturb} conditions praised the relevance of the suggestions. For example, one user wrote: {\it ``I appreciated the feedback, it gave me specific points to elaborate or improve on.''} These comments illustrate the promise of selecting feedback with Perturbation Analysis. Moreover, 3 users in {\it Seg+Perturb} condition praised the localized feedback. One wrote: {\it ``I felt like the segment-level feedback was very informative compared to the document-level feedback. I thought it was very easy to read.''} However, no users in the {\it Seg+Krause} condition praised the localized feedback, demonstrating that localized feedback must also be relevant to be useful. 

\sstitle{User Concerns:}
A general concern (N=8) was the lack of trust for computer-generated feedback; one user stated that they didn't make changes because {\it ``It seemed like the feedback was just an automated response...''}  We speculate that interface design and personalized feedback text, which indicate that the feedback is customized, may help with this problem.  For instance, this could be done by leveraging the conversation tones from chat interfaces~\cite{fast2017iris,lasecki2013chorus}.

6 users criticized the {\it Perturb} condition because pressing {\it Recompute Text Feedback} after editing their writing produced the same, or a subset, of the initial feedback.   One user wrote: {\it ``Initially it seemed interesting but even after editing... nothing changed when I resubmitted.''}  This was understandable since not all edits necessarily reclassify the text, however, it highlights the need to take editing {\it changes} and prior feedback into account.

%

\section{Limitations and Future Work}
There are many ways to improve upon our findings.  First, is to continue improving quality prediction models and to develop new writing features for different communities.  This directly improves the quality of our generated feedback.  Second, is to enrich the feedback text generation to account for personalization, past feedback, and nuances in the perturbations. Third, is to integrate existing feedback systems such as Grammarly~\cite{grammarly} that are specialized to address a specific writing characteristic such as readability or grammar errors.

There are several limitations to address in future work.   First, the heuristic that we present for Perturbation Analysis is limited to tree-based classifiers.  An important direction for future work is to extend Perturbation Analysis beyond tree-ensembles to other linear and kernel-based models, as well as increasingly popular deep learning models.   Our problem formulation in Section~\ref{s:perturb} formalizes this general problem, but developing efficient techniques is still challenging. Recent work on surrogate model explanations~\cite{ribeiro2016model} and adversarial generation~\cite{rajeswar2017adversarial} are promising approaches to adopt.  

Second, privacy limits the features for which Perturbation Analysis can provide feedback.  Many models rely on personally sensitive features such as location, interests, or past behavior that may either be unavailable in a real deployment scenario or unrelated to the writing content.  For future systems to be successful in practice, they must use models trained only on relevant, non-sensitive writing features. We show that this is possible because our evaluation uses only features extracted directly from the users' writing.

Third, unlike Perturbation Analysis, segment-level feedback only showed statistically significant improvements for the Amazon when combined with Perturbation Analysis.  We suspect that segment-based feedback may be most useful when documents contain many segments, and when the feedback itself is relevant.   Further, our weak supervision approach may not generalize to settings where document quality does not correlate with segment quality.  These settings would require crowdsourced labels for the segments or techniques to infer their labels~\cite{kotzias2015group}.  Finally, the best feedback may not be to treat each segment independently, and techniques to account for dependencies between segments can further improve feedback quality.

\section{Conclusion}

This paper combines three emerging and active research areas---text quality prediction models, model explanation, and text segmentation---to generate localized writing feedback that is cognizant of community quality standards.  Our Perturbation Analysis technique analyzes a pre-trained community-specific quality prediction model to recommend combinations of changes to writing features which will most improve the predicted quality. Perturbation Analysis consistently outperforms a baseline feedback generation method. We also showed that providing localized feedback, via a weak-supervision approach that uses document-level labels to label individual segments, is promising when combined with Perturbation Analysis.

\section{Acknowledgements}

Special thanks to James Sands, Kevin Lin, and Rahul Khanna for engineering contributions to the system and interface; Jiannan Wang, Lydia Chilton, and Eric Riesel for assistance in writing and framing our contributions; Markus Krause, Xiao Ma, Mor Naaman, and Julian McAuley for datasets and baselines; Niloufar Salehi, Robert Netzorg, Ali Alkhatib, and Philippe Cudre-Mauroux for valuable feedback. This work was supported by an Amazon Research Award; NSF \#1527765 and its REU.

 {
 \fontsize{9.0pt}{10.0pt}
 \selectfont
 \bibliography{Nilforoshan}
 \bibliographystyle{aaai}
 }

\clearpage

\section{Appendix}

\subsection{Perturbation Heuristic for Tree Ensembles} 

The space of solutions for Problem~\ref{p:problem} is exponential in the number of features used by the model, because the cardinality of the power set $|\mathcal{P(\mathcal{F})}| = 2^{|\mathcal{F}|}$, meaning that for $n$ features there are $2^{n}$ possible sets of perturbations to naively explore. 
We instead present a heuristic solution whose complexity is linear in the number of paths in the random forest model.
The main idea is to scan each tree in the random forest and compute perturbations and scores local to the tree.

Let the $D=\{d_1,\dots,d_m\}$ be the training dataset and $Y=\{y_1,\dots,y_m\}$ be their labels.
The ensemble tree model $M=\{T_1,\dots,T_t\}$ is composed of a set of trees.
A tree $T_i$ is composed of a set of $k$ decision paths $q_i^1,\dots,q_i^k$; each path $q_i^j$ matches a subset of the training dataset $D_i^j\subseteq D$ and its vote $v_i^j$ is the majority label in $D_i^j$.
Thus, the output of $T_i(d)$ is the vote of the path that matches $d$, and the output of the ensemble $M(d)$ is the majority vote of its trees.

Let $minp(d, q_i^m)$ return the minimum perturbation $p$ (based on its L2 norm) such that $d$ matches path $q_i^m$:
$$minp(d, q_i^m) = \argmin_{p \in \mathbb{R}^n} |p|_2\ s.t.\ q_i^m\ \textrm{matches}\ d+p$$
Rather than examining all possible perturbations, our heuristic to compute $S^d_i$ restricts the set of perturbations with respect to the decision paths in the trees that increase $d$'s quality (e.g., $p^1$ and $p^2$ in Figure~\ref{f:tree}).  The impact function $I()$ is identical, however it takes a path $q_i^j$ as input and internally computes the minimum perturbation $minp(d,q_i^j)$.  Finally, we compute the confidence $C(d)$ as the fraction of samples in $D_i^j$ whose labels $y_k$ match the path's prediction $v_i^j$.    
{\small
\begin{align*}
S^d_i &= \sum_{T_i \in M} \sum_{q_i^j \in T_i} I(d, q_i^j)  \ \textrm{if}\  v_i^j > M(d)\\
I(d, q_i^j)   &= \frac{v_i^j - M(d)}{\Delta(minp(d, q_i^j))}\times C(d+minp(d, q_i^j))\\
C(d) &= \frac{|\{d_k \in D_i^j | y_k = v_i^j \}|}{|D_i^j|}
\end{align*}
}

Given $d$ and predicted quality $M(d)$, we retrieve and scan the paths with higher quality.
For each scanned path $q$, we compute the change in the quality function, discount its value by the minimum perturbation $p$ and the path's confidence.  
Finally, we add this value to the score of all features perturbed in $p$.

\subsection{Crowdsourcing Segment Labels}

\stitle{Procedure: }
We filtered the Amazon review corpus to laptop reviews that had received at least 15 or more votes, and defined helpful reviews as those with $>60\%$ helpful votes.  We then ran these reviews through the TopicTiling~\cite{riedl2012topictiling} to build a corpus of review segments. We then selected 250 random segments from helpful and unhelpful reviews each (500 total).
We restricted participation to workers with 95\% HIT approval rate, 75\% HIT submission rate, and US residents. Each task, including training and qualification was $\$0.04$, and we paid a  $\$1.00$ bonus for labeling all 50 segments.

\begin{figure}
\centering
\includegraphics[width=1.0\columnwidth]{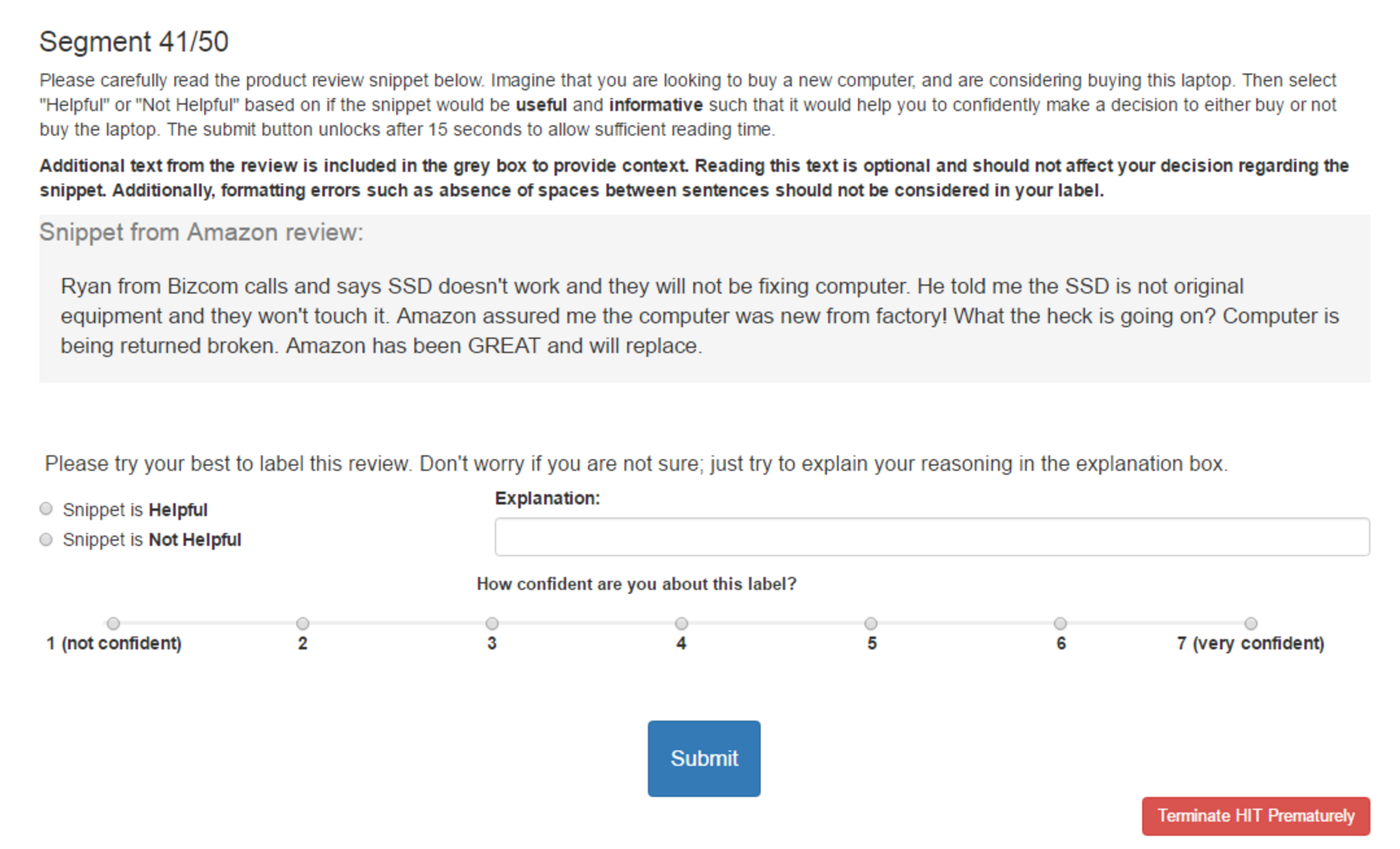}
\caption{Example of the segment helpfulness labeling task interface.  Participants read a review segment (grey box) and label it as helpful or unhelpful, state their prediction confidence, and explanation the rationale.} 
\label{f:userstudy_seg}
\end{figure}

Participants were asked to label the helpfulness of up to $50$ segments (task interface shown in Figure~\ref{f:userstudy_seg}). 
The training phase showed a guide with three examples each of helpful and unhelpful segments, along with a brief explanation of the rationale for the label.  Then, participants labeled $18$ training segments, and were shown an explanation after submission.  All training segments were hand-selected to span a variety of sentiments, lengths, topics, and helpfulness ratings.  The explanations were decided by consensus among the lead authors.  The qualification task consisted of three segments that were clearly helpful or unhelpful.  Finally, participants were shown $50$ random segments and asked to label them as helpful or unhelpful. Participants were also asked to provide their confidence and a brief explanation of their decisions, though these data were not used in determining a segment's label. Participants were given the option to end the task and submit their work at any point.

\stitle{Results:} 
There were $67$ participants, and each labeled on average $23.2$ segments. We ran the experiment until each segment received $\ge3$ votes. We used Get Another Label~\cite{sheng2008get} to determine the helpfulness of each segment. These labels allow us to evaluate the segment-level classifier, trained using model features from (Section 5.1). We computed pairwise accuracies between the document labels, classifier predictions, and crowd labels: $71.1\%$ (classifier predicting crowd label), $72.5\%$ (classifier predicting Document label), and $69.5\%$ (document label predicting crowd label). The consistent results between all three comparisons suggest the efficacy of the segment-level classifier, and our end-to-end experimental results suggest that the predictive model is effective at providing segment level feedback. Nevertheless, more studies are needed to fully evaluate this hypothesis across other text domains and document lengths. We defer this to future work.

\end{document}